\documentclass[aps,prl,twocolumn,superscriptaddress]{revtex4-1}

\usepackage[utf8]{inputenc}
\usepackage{chemformula}
\usepackage{amsmath}
\usepackage{mathrsfs}
\usepackage{multirow}
\usepackage{amssymb}
\usepackage{graphicx}
\usepackage{siunitx}
\usepackage{lipsum}
\usepackage{colortbl}
\usepackage{diagbox}
\usepackage{booktabs}
\usepackage{hhline}

\usepackage[normalem]{ulem}

\newcommand{\angstrom}{\textup{\AA}}
\usepackage[colorlinks=true, allcolors=blue]{hyperref}
\usepackage{xr}
\makeatletter
\newcommand*{\addFileDependency}[1]{
  \typeout{(#1)}
  \@addtofilelist{#1}
  \IfFileExists{#1}{}{\typeout{No file #1.}}
}
\makeatother

\definecolor{mag}{RGB}{255,0,255}

\begin{document}
\title[Data-Augmentation for Graph Neural Network Learning of the Relaxed Energies of Unrelaxed Structures]{Data-Augmentation for Graph Neural Network Learning of the Relaxed Energies of Unrelaxed Structures }

\author{Jason B. Gibson}
\affiliation{Department of Materials Science and  Engineering, University of Florida, Gainesville, Florida 32611, USA}
\affiliation{Quantum Theory Project, University of Florida, Gainesville, Florida 32611, USA}
\author{Ajinkya C. Hire}
\affiliation{Department of Materials Science and  Engineering, University of Florida, Gainesville, Florida 32611, USA}
\affiliation{Quantum Theory Project, University of Florida, Gainesville, Florida 32611, USA}
\author{Richard G. Hennig}
\email{rhennig@ufl.edu}
\affiliation{Department of Materials Science and  Engineering, University of Florida, Gainesville, Florida 32611, USA}
\affiliation{Quantum Theory Project, University of Florida, Gainesville, Florida 32611, USA}

\date{\today}

\begin{abstract}
 Computational materials discovery has continually grown in utility over the past decade due to advances in computing power and crystal structure prediction algorithms (CSPA). However, the computational cost of the \textit{ab initio} calculations required by CSPA limits its utility to small unit cells, reducing the compositional and structural space the algorithms can explore. Past studies have bypassed many unneeded \textit{ab initio} calculations by utilizing machine learning methods to predict formation energy and determine the stability of a material. Specifically, graph neural networks display high fidelity in predicting formation energy. Traditionally graph neural networks are trained on large data sets of relaxed structures. Unfortunately, the geometries of unrelaxed candidate structures produced by CSPA often deviate from the relaxed state, which leads to poor predictions hindering the model's ability to filter energetically unfavorable prior to \textit{ab initio} evaluation. This work shows that the prediction error on relaxed structures reduces as training progresses, while the prediction error on unrelaxed structures increases, suggesting an inverse correlation between relaxed and unrelaxed structure prediction accuracy. To remedy this behavior, we propose a simple, physically motivated, computationally cheap perturbation technique that augments training data to improve predictions on unrelaxed structures dramatically. On our test set consisting of 623 Nb-Sr-H hydride structures, we found that training a crystal graph convolutional neural networks, utilizing our augmentation method, reduced the MAE of formation energy prediction by 66\% compared to training with only relaxed structures. We then show how this error reduction can accelerates CSPA by improving the model's ability to filter out energetically unfavorable structures accurately.
\end{abstract}

\maketitle

\section{Introduction}

The discovery of novel functional materials drives innovation. The process of discovery has dramatically accelerated over the past decade, partially as a product of growing crystal structure databases~\cite{Jain2013, Kirklin2015, draxl_scheffler_2018, CURTAROLO2012218} and improved computationally based CSPA~\cite{oganov_pickard_zhu_needs_2019}, such as genetic algorithms (GA)~\cite{GASP-Python}, basin hoping~\cite{BasinHop}, elemental substitution~\cite{MgMnOData}, and particle swarm techniques~\cite{ParticleSwarm}. CSPA have played a prominent role in successfully predicting the structure and establishing the stability of high-pressure, high-temperature superconducting binary hydrides~\cite{Duan2014,Liu2017,Peng2017}. Many recent studies have started looking for stable ternary hydride superconductors. The addition of a third element to the binary hydrides can potentially stabilize these materials at much lower pressure\cite{Sun2019,DiCataldo2021,Hilleke2022}. Complex ternary and quaternary materials systems are also promising candidates for hydrogen storage applications\cite{Huang2021}. 

In particular, GAs have proven their utility to identify thermodynamically stable phases efficiently; successfully identifying novel materials for applications such as Li-Ge batteries~\cite{Tipton2014133} and solar cells~\cite{PhysRevLett.111.165502}. Unfortunately, finding thermodynamically stable phases in ternary and quaternary systems is notoriously difficult due in part to the computationally expensive \textit{ab initio} calculations required to relax and calculate the energies of GA produced structures, accounting for 99\% of the algorithm's computational cost.~\cite{Heiles} This places restrictions on the size and the composition of the unit cells, inhibiting the exploration of complex material systems.

This computational cost can be reduced by bypassing many of the costly \textit{ab initio} calculations via a machine-learned filter or by implementing a more computationally efficient machine-learned surrogate potential to pre-relax structures~\cite{xie2021ultrafast}.
Wu {\it et al.}~\cite{Wu2013} fitted a classical potential to the structures evaluated by density functional theory (DFT). They used the potential to pre-relax the structures in the GA and only evaluated the best structures with DFT. Jennings {\it et al.}~\cite{Jennings2019} used a machine learning (ML) model to predict a structure's fitness directly and then only used DFT to evaluate structures that improved the current population. These methods are still somewhat hindered because many DFT evaluated structures are required to train a ML model to an adequate fidelity. Further, the models are specific to the materials' space the GA is searching, restricting their application to the given GA search.

Alternatively, there has been work to create universal ML models that determine a material's stability by predicting the formation energy of structures containing elements across the periodic table. Most notably, Xie {\it et al.}~\cite{Xie2018} predicted formation energy using a crystal graph convolutional neural network (CGCNN) trained on the materials project (MP) database~\cite{Jain2013}. The CGCNN represents a crystal structure as a multi-graph and builds a graph convolutional neural network on top of the multi-graph. This enables the model to learn the best features to represent the structure as opposed to the typical "handcrafted" feature approach~\cite{Zeeshan} and achieve a formation energy validation MAE of 39~meV/atom~\cite{Xie2018}. More recently, the MAE of formation energy prediction of graph-based models continued to decrease to 21-39~meV/atom~\cite{choudhary_decost_2021, GeoCGNN, MegNet, Park, NOH}.

However, Park {\it et al.}~\cite{Park} found that the model's dependence on a structure's atomic coordinates hinders the model's predictive fidelity on structures that strongly deviate from their relaxed states. Given that to obtain a structure in a relaxed state, a DFT relaxation and hence energy calculation is needed, the reported MAEs do not represent the model's ability to accurately identify unrelaxed structures that would relax to stable structures. On an unseen test set of 311 stable ThCr$_2$Si$_2$-type compounds, the CGCNN obtained a reasonable formation energy MAE on relaxed structures of 56~meV/atom. However, the prediction MAE was 370 meV/atom for the same test set on unrelaxed structures. This high error led to a true positive rate (TPR) of 0.48 when filtering the unrelaxed compounds in the data set~\cite{Park}.

Noh {\it et al.}~\cite{NOH} directly addressed the high formation energy prediction MAE of unrelaxed structures by adding two forms of regularization to the CGCNN in their CGCNN-HD approach. Replacing the softplus activation function in the convolution function with a hyperbolic tangent and adding dropout layers~\cite{JMLR:v15:srivastava14a} between each fully connected layer, reduced the MAE of the formation energy for a test set of unrelaxed Mg-Mn-O compounds from 518~meV/atom to 296~meV/atom.

\begin{figure}[tb]
    \includegraphics[width=.48\textwidth]{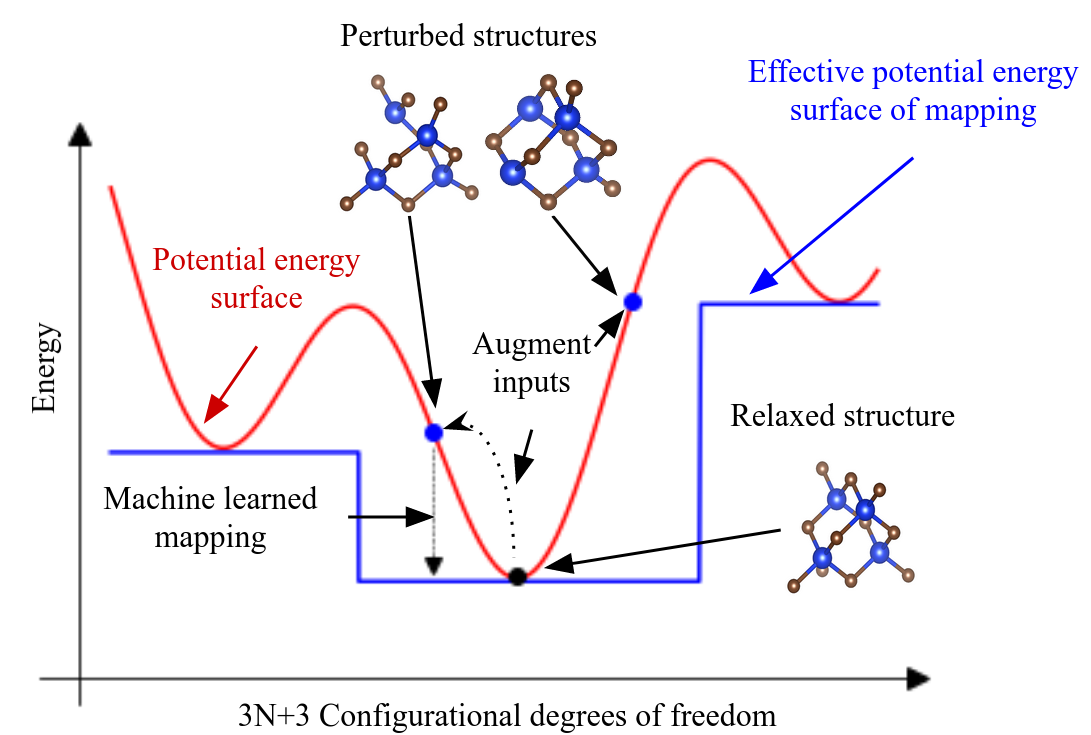}
    \caption{Data augmentation for learning the potential energy surface (PES). The red line denotes a 2D representation of the continuous PES of materials. The blue line illustrates the effective PES, which describes the energy of a relaxed structure for a given unrelaxed input structure. Data augmentation aims to improve the machine learning of this effective PES by better sampling the configuration space. The black circle indicates the relaxed structures contained in the data set, and the blue circles symbolize artificially generated structures for the data augmentation.}
    \label{PEL_as_step}
\end{figure}

The significant errors for unrelaxed structures are due to the limited sampling of the complex multi-dimensional configuration space of the potential energy surface (PES), with the relaxed structures only describing the minima of the surface. Since unrelaxed structures are not located at these minima, predicting a structure's formation energy at an unrelaxed configuration is a qualitatively different task~\cite{goodall2021rapid}. To sample configurations near the minima of the PES, Smith {\it et al.}~\cite{Smith2017} applied a data augmentation technique known as normal mode sampling to a large dataset of molecular structures, resulting in a high fidelity neural network potential. However, determining the normal modes requires millions of phonon calculations which, while attainable for molecular structures, is infeasible for crystal structures.

Recently, Honrao {\it et al.}~\cite{Honrao2020} showed that GA data could be used to predict relaxed formation energies of unrelaxed structures to high accuracy. This high accuracy was achieved by augmenting the training data set, setting the formation energy of every structure within a basin of attraction to the minima of the respective basin of attraction, essentially modeling the continuous PES as a step function.

We propose leveraging these findings to improve formation energy prediction of unrelaxed structures by augmenting our data using a simple, physically motivated perturbation technique. Figure~\ref{PEL_as_step} illustrates our augmentation approach, which perturbs the atomic coordinates of a relaxed structure to generate additional training points that describe the regions surrounding the minima of the PES. We then map these perturbed structures to the energy of the relaxed structure, which requires no additional \textit{ab initio} calculations. We utilize the CGCNN~\cite{Xie2018} and CGCNN-HD~\cite{NOH} to analyze how the augmentation affects formation energy predictions. We train these models on the MP database~\cite{Jain2013} augmented by perturbed structures. The resulting CGCNN models have similar prediction errors as the original ones for relaxed structures. To show the improvement in formation energy prediction, we apply the models to a test set consisting of 623 unrelaxed Nb-Sr-H hydride structures produced from a GA structure search. We find that compared to training on only relaxed structures, training with the augmented data set, consisting of relaxed and perturbed structures, reduced the formation energy prediction MAE from 251~meV/atom to 86~meV/atom for CGCNN and from 172~meV/atom to 82~meV/atom for CGCNN-HD, as compared to the models trained only on relaxed structure.

\section{Methods}
For training, we use two data sets derived from the MP database~\cite{Jain2013} accessed on December 10, 2021. The 1\textsuperscript{st} data set, referred to as the relaxed data set, consists of 126k relaxed structures from the MP database, 20\% of this data is held out for validation. The 2\textsuperscript{nd} data set, referred to as the augmented data set, consists of the relaxed set and one perturbed structure for every relaxed structure.

\begin{figure}[tb]
    \includegraphics[width=.45\textwidth]{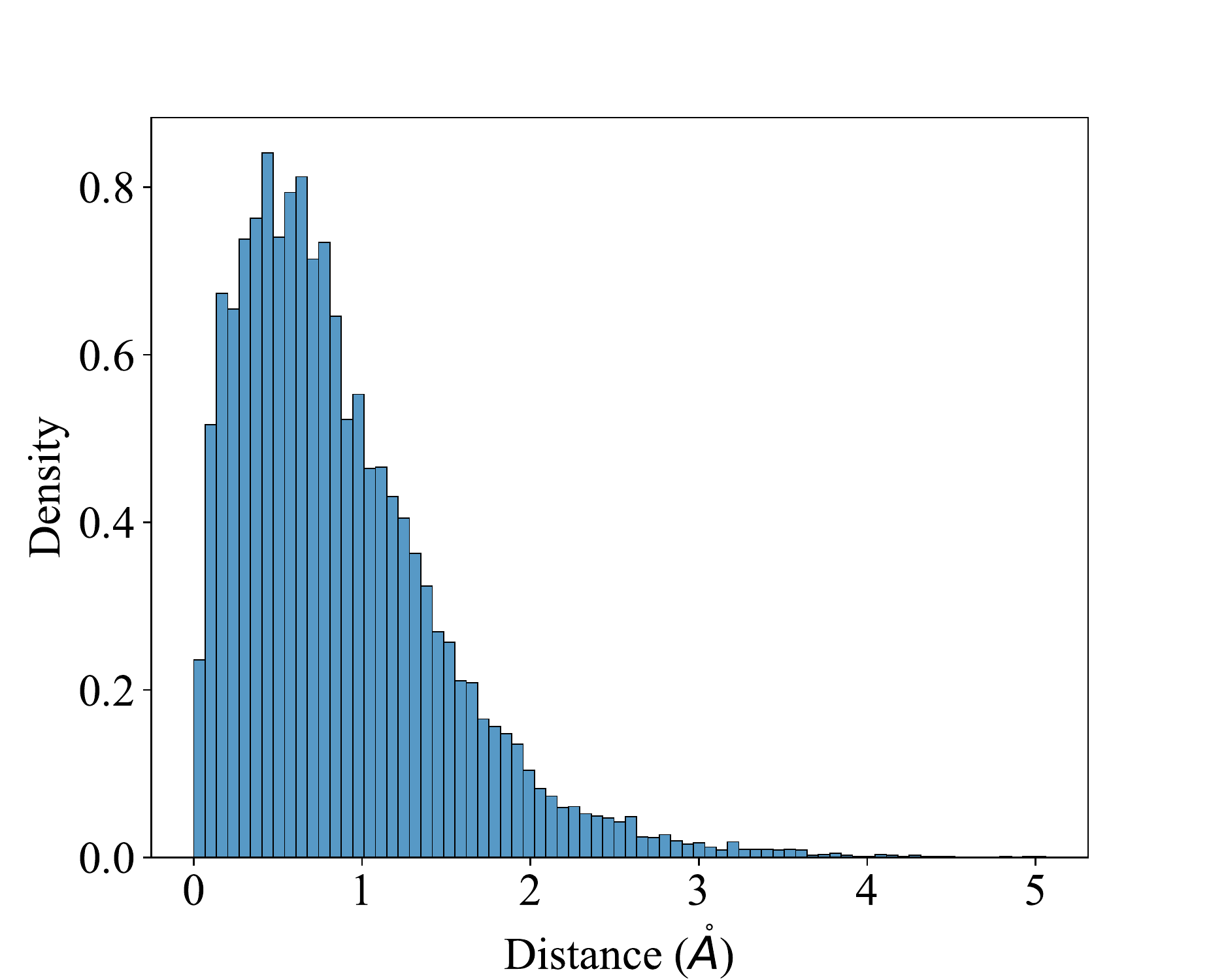}
    \caption{Distribution of the displacement of atoms during structural relaxation in three separate GA runs. The displacement is measured by the change in fractional coordinates multiplied by the lattice vector matrix of the relaxed structure.}
    \label{dist_hist}
\end{figure}

We augment the data by perturbing the coordinates, $R_i$ of all atoms,  $i$, in each relaxed structures using a displacement vector 

\begin{equation} \label{eq:1}
    \Delta \vec{R}_{i} = (M_{i}^x, M_{i}^y, M_{i}^z)
\end{equation}
where each Cartesian component is obtained by sampling a displacement distribution. The displacement distribution was determined by analyzing the displacements of atoms during relaxation in three separate GA structure searches. The details of the GA structure searches can be found in the supplemental materials.
The distance was determined by first taking the difference between the initial and final structure's fractional coordinates using the minimum image convention~\cite{Deiters+2013+345+352}. 
These differences were then multiplied by the lattice vector matrix of the relaxed structure to obtain the cartesian displacement vector and the euclidean norm for this vector. Figure~\ref{dist_hist} shows the resulting distribution of displacements, which was then fitted to a Gaussian mixture model (GMM) as implemented in the open python library scikit-learn~\cite{Scikit}. The value of $M_{pi}$ was then selected by randomly sampling the GMM. Information about the change in lattice vectors and volume can be found in Figure~S3. 

The CGCNN and CGCNN-HD are trained on both the relaxed and augmented data. The model's architecture was determined by performing a grid search on the CGCNN trained on the augmented data. Table~S2 provides the range of parameters considered in the grid search. The architecture that minimized the validation error consists of 3-graph convolutional layers followed by 6-hidden layers with 64 neurons each. This architecture was then used for all models. Interestingly we found that models with more than eight hidden layers suffered from the vanishing gradient problem~\cite{vangradient} when training on the relaxed data while training on the augmented data allowed for deeper models. The remaining model hyperparameters are set to the values reported in Ref.~\cite{Xie2018} for CGCNN and Ref.~\cite{NOH} for the CGCNN-HD.

\begin{figure}[t]
    \centering
    \includegraphics[width=\columnwidth]{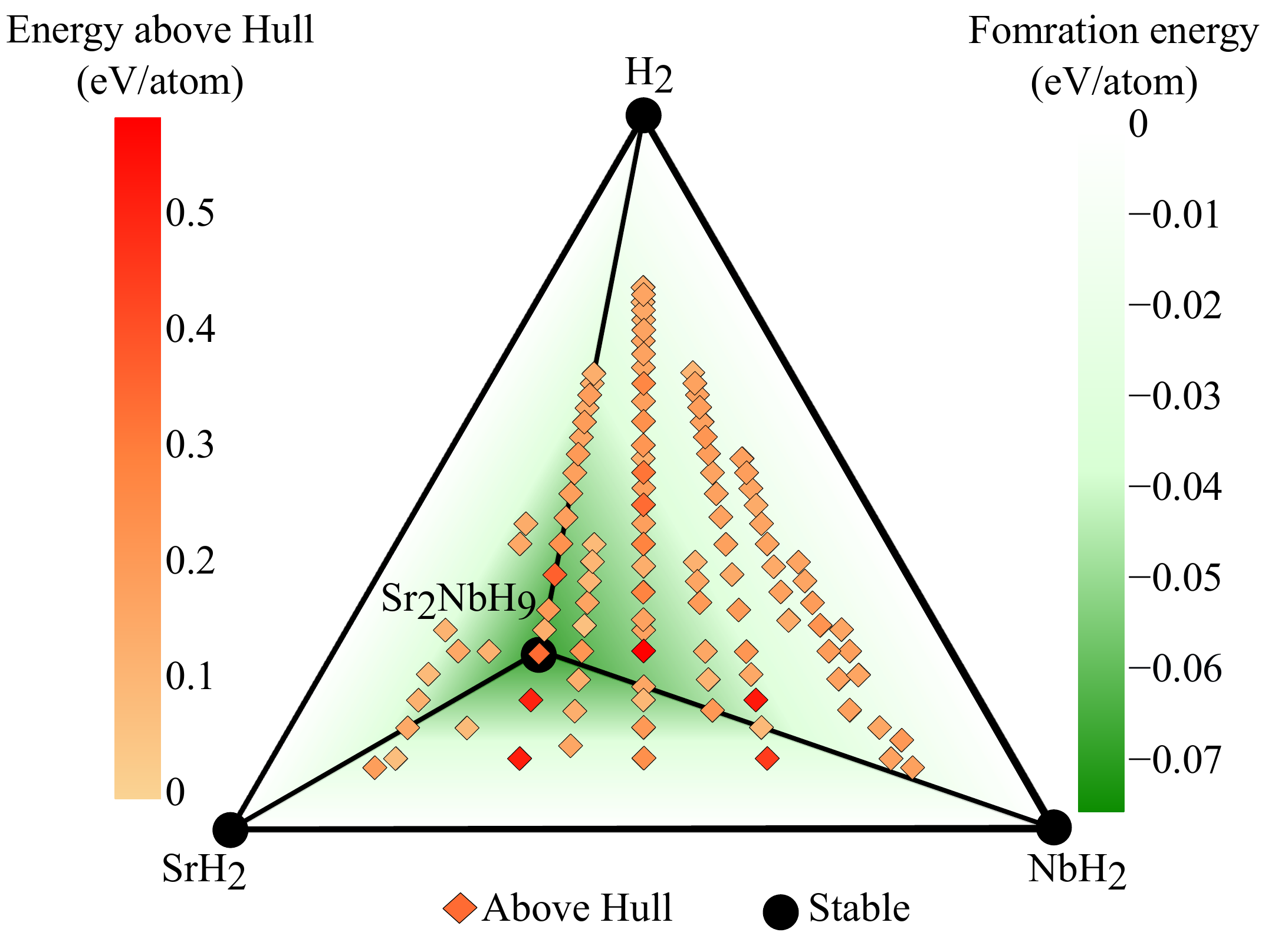}
    \caption{The convex hull of the energies of the predicted structures for the NbH$_2$-SrH$_2$-H$_2$ materials system. The distance from the convex hull measures the thermodynamics stability of the various candidate compounds. The green shading indicates the formation energy of the thermodynamically stable compounds and mixtures relative to the three compounds NbH$_2$, SrH$_2$, and H$_2$. We identify a previously unknown ternary hydride, Sr$_2$NbH$_9$.}
    \label{hydride_convex_hull}
\end{figure}

To provide test data for our models, we performed a GA search over the ternary system formed by H$_2$-Sr$_6$NbH$_{16}$-Nb$_6$SrH$_{16}$. We used the Genetic Algorithm for Structure and Phase Prediction (GASP) python package~\cite{Tipton_2013,revard2016grand} for performing the GA search. Our aim with the search was to produce high hydrogen-containing structures that might show superconductivity. 
We remove the elemental hydrogen structures and partitioned this data into two test sets consisting of the relaxed and unrelaxed hydrides, referred to as ``H-Relaxed'' and ``H-Unrelaxed,'' respectively. Additionally. since the MP database contains few hydride structures, this data provides a challenging test case.

\begin{figure*}[ht]
    \centering
    \includegraphics[width=\textwidth]{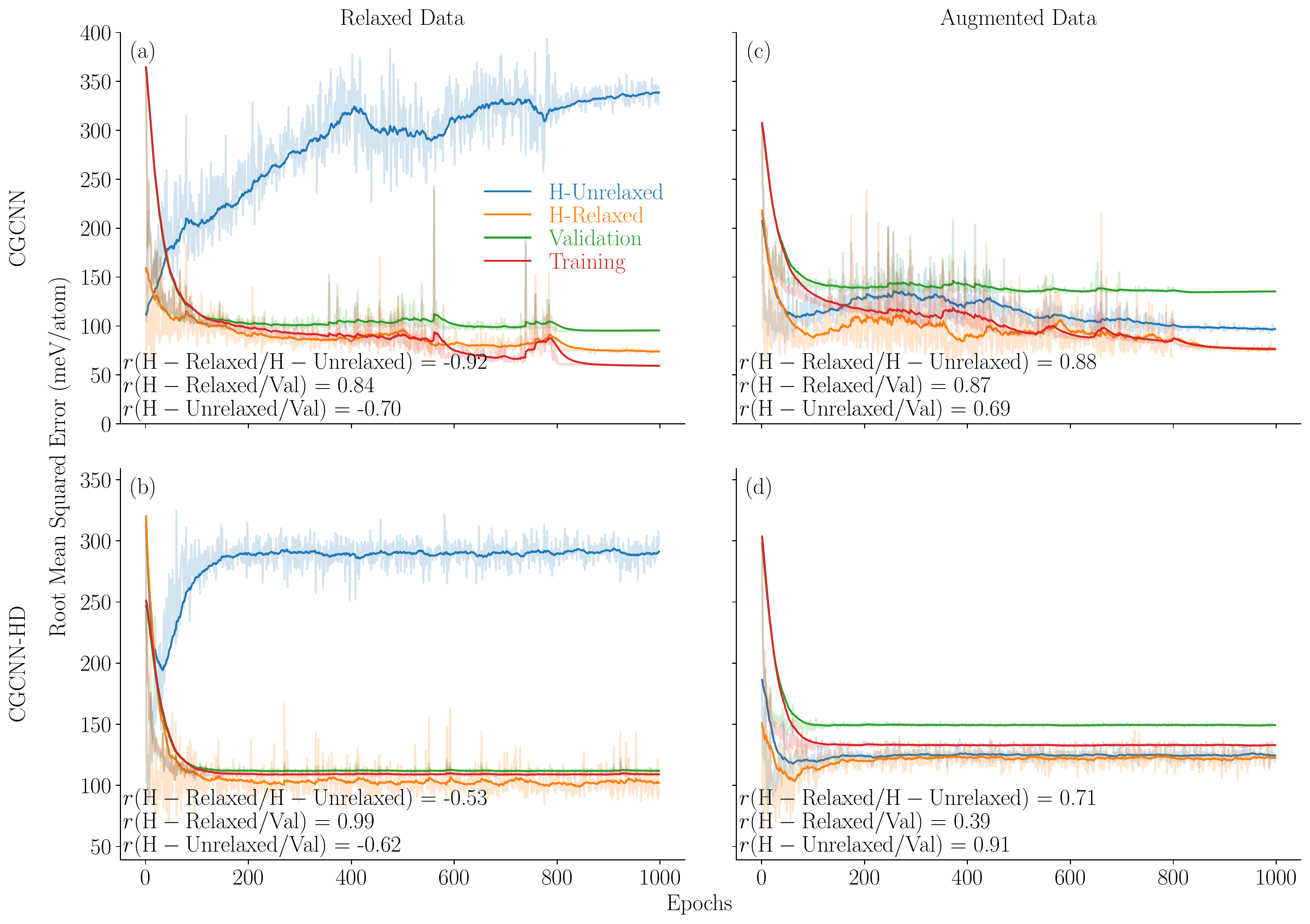}
    \caption{Learning curves for the CGCNNN (1\textsuperscript{st} row) and CGCNN-HD (2\textsuperscript{nd} row) trained on the relaxed  (1\textsuperscript{st} column) and augmented (2\textsuperscript{nd} column) data. The faded curves show the exact loss values, while the solid curves show the smoothed values. The red and green curves denote the loss on the training and validation data, respectively. The orange and blue curves display the loss for the H-Relaxed and H-Unrelaxed test sets, respectively. Note that the H-Relaxed and H-Unrelaxed data sets were not used in the training or validation of the model. The $r$ values are the Pearson correlation coefficients between the stated trends.}
    \label{np_lc}
\end{figure*}

To relax and evaluate the energies of the candidate structures generated by GASP, we use VASP~\cite{PhysRevB.47.558,PhysRevB.49.14251,KRESSE199615,PhysRevB.54.11169} with the projector augmented wave method~\cite{PhysRevB.50.17953} and the Perdew-Burke-Ernzerhof (PBE) generalized gradient approximation for the exchange-correlation functional~\cite{PhysRevLett.77.3865}. We use a $k$-point density of 40 per inverse \angstrom~with the Methfessel-Paxton scheme and a smearing of 100~meV for the Brillouin zone integration, and a cutoff energy of 250~eV for the plane-wave basis set. The GA search was terminated after 771 DFT relaxations. We recomputed all energies using the VASP inputs generated by the MPRelaxset class of pymatgen to ensure consistency between the training and test sets and computed the formation energies~\cite{Wang2021}. Figure~\ref{hydride_convex_hull} shows the ternary convex hull of the Nb-Sr-H system produced using the GA-generated structure and the known competing phases from the MP database. Noteworthy, our structure search found a new, previously unreported, thermodynamically stable ternary hydride, Sr$_2$NbH$_9$. Preliminary analysis of Sr$_2$NbH$_9$ suggest that the band gap closes at around 100 GPa.

\section{Results} \label{Results}

Figure~\ref{np_lc} shows the training, validation, and test (H-Relaxed/H-Unrelaxed) RMSE for each training epoch of the respective models. For interpretability, the trends are smoothed using an exponential moving average with a smoothing weight of 0.95. The Pearson correlation coefficients~\cite{freedman2007statistics}, with a value of 1 for perfect correlation and -1 for perfect anti-correlation, are computed between smoothed trends of H-Relaxed/H-Unrelaxed, H-Relaxed/validation, and H-Unrelaxed/validation. 

\begin{figure*}[t!]
    \centering
    \includegraphics[width=\textwidth]{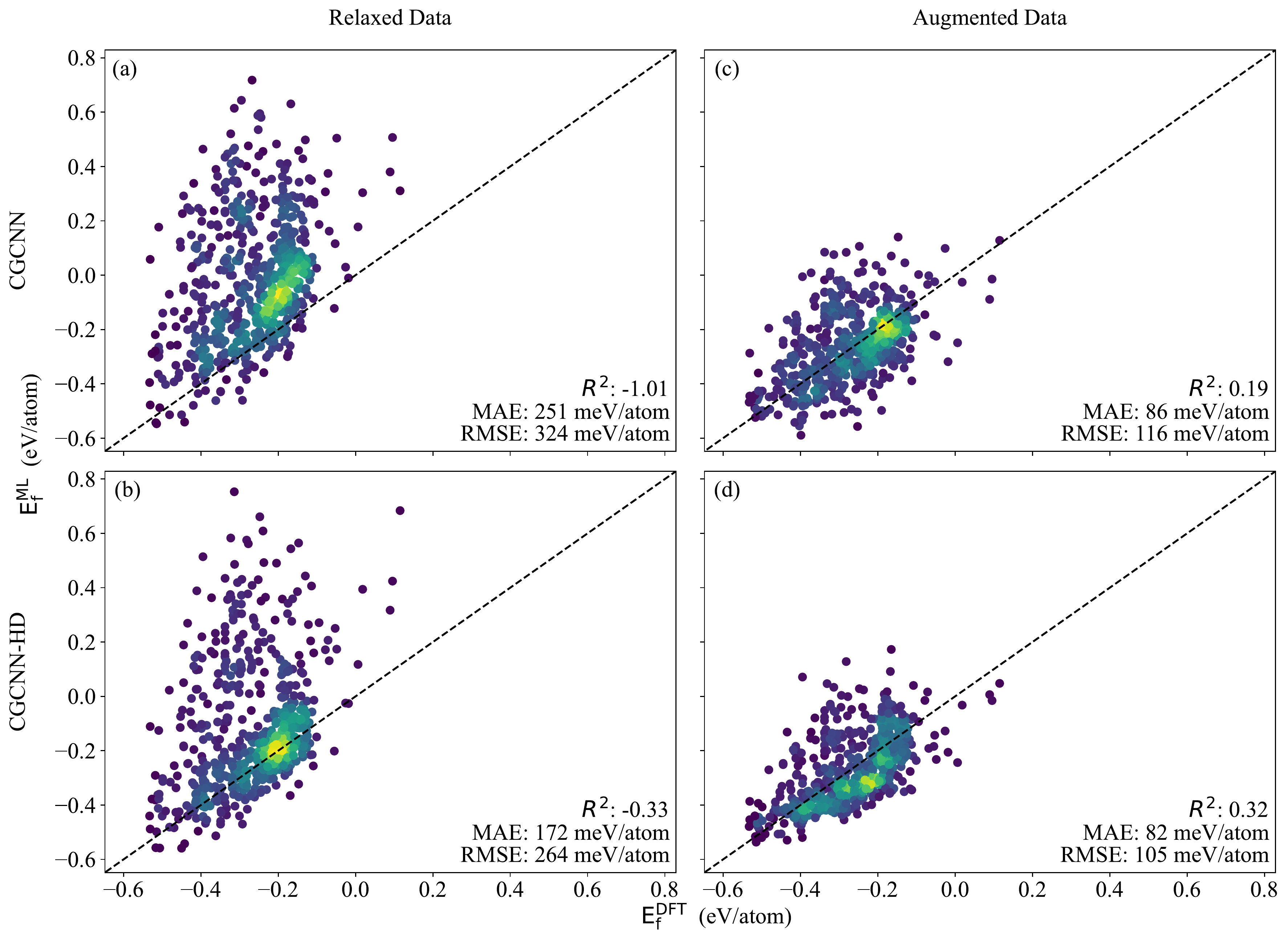}
    \caption{The Parity plots for the CGCNNN (1\textsuperscript{st} row) and CGCNN-HD (2\textsuperscript{nd} row) trained on the relaxed  (1\textsuperscript{st} column) and augmented (2\textsuperscript{nd} column) data. The $x$-axis denotes the DFT computed formation energies, while the $y$-axis denotes the predicted formation energies. The values reported in the lower right are the coefficient of determination, $R^2$, the MAE, and the RMSE.}
    \label{bests}
\end{figure*}

The Pearson correlation coefficients show that the CGCNN and CGCNN-HD trained on only relaxed structures (Figures~\ref{np_lc}(a) and (b)) result in anti-correlated trends between predictions on the relaxed and unrelaxed structures of the test set. These findings demonstrate that accurate predictions on relaxed structures do not lead to accurate predictions on unrelaxed structures and provide insight into the high prediction error for unrelaxed structure inputs reported in the literature. Furthermore, the CGCNN and CGCNN-HD trained on only relaxed structures display anti-correlated trends between the H-Unrelaxed test set and the validation set. This anti-correlation is detrimental to the model's predictive performance on unrelaxed structures because the validation error shows that predictions are improving and training should continue, while in actuality, unrelaxed structure predictions are getting worse. Additionally, the anti-correlation leads to an inability to correctly optimize a model's hyperparameters.

Figures~\ref{np_lc}(c) and (d) illustrate the effectiveness of training with the augmented dataset. Simply perturbing the atomic coordinates of each relaxed structure structure and then training on both the relaxed and perturbed structures dramatically improved the models' predictive ability on unrelaxed structures. Both the CGCNN and CGCNN-HD trained on the augmented data set show a high Pearson correlation between H-Unrelaxed/validation, enabling the effective optimization of a model's hyperparameters and the implementation of early stopping. 

Figure~\ref{bests} compares the models' formation energy predictions on the unrelaxed hydride test data to the DFT-computed formation energies. The CGCNN in Figure~\ref{bests} (a) trained on only relaxed data tends to over predict $\mathrm{E_{f}}$ for the higher energy hydrides, which leads to a significant prediction MAE of 251~meV/atom. The added regularization applied to the CGCNN-HD in Figure~\ref{bests} (b) improves the predictions on the higher energy hydrides. Still, the model tends to over predict $\mathrm{E_{f}}$ leading to an MAE of 172~meV/atom when training on relaxed.

Training with the augmented dataset substantially improves the prediction MAEs for both the CGCNN and CGCNN-HD, reducing the prediction MAE to 86~meV/atom and 82~meV/atom, respectively. While the CGCNN-HD trained on augmented data has the lowest testing MAE, the dense region of underpredicted formation energies seen in Figure~\ref{bests}~(d)  leads to a substantial number of misclassified unstable structures, hindering the model's ability to filter unstable structures. This will be discussed further in the proceeding section.

Due to the large perturbations, many augmented structures have likely moved to neighbor basins of attraction. Intuitively this would seem to yield substantial errors. However, as shown previously, the prediction of unrelaxed structures improved substantially. We suspect the error associated with perturbing a structure to a neighboring basin of attraction is mitigated due to the tendency of neighboring basins to cluster around similar minima on the PES~\cite{Revard2014}. Still, the large perturbations likely introduce some prediction error. A more sophisticated augmentation method could likely reduce the number of structures perturbed into neighboring basins and further improve predictions. 

Interestingly, the models trained on the augmented data also display improved predictions on H-Relaxed (Figure~S1). However, this improvement seems to be specific to our test data as predictions on the relaxed validation data were better when the model was trained only on relaxed structures(Figure~S2). The CGCNN-HD trained on only relaxed structures under predicted the structures in H-Relaxed likely because the training data contains relatively few transition metal hydrides and the bounded hyperbolic activation function, utilized in the CGCNN-HD's convolutional layers, has poor predictive power on unseen domains~\cite{Kim2021}. This poor predictive power on unseen domains is also the reason the CGCNN-HD models make poor predictions on the the high and low formation energy structures of the MP data.

\subsection{Filtering unstable hydrides}
To evaluate the models' ability to filter energetically unfavorable structures, we removed the MP database's correction applied to hydrogen-containing compounds and constructed a convex hull using the five known competing phases of the Nb-Sr-H system. Then, based on this constructed convex hull, we computed the hull distance $\mathrm{E_{Hull}^{DFT}}$of all the structures in the test set, utilizing their DFT-computed formation energy, and defined all structures with $\mathrm{E_{Hull}^{DFT} < 0}$ as stable. A total of 10 structures in the test set met the stability criteria.

\begin{figure}[htbp]
    \includegraphics[width=.48\textwidth]{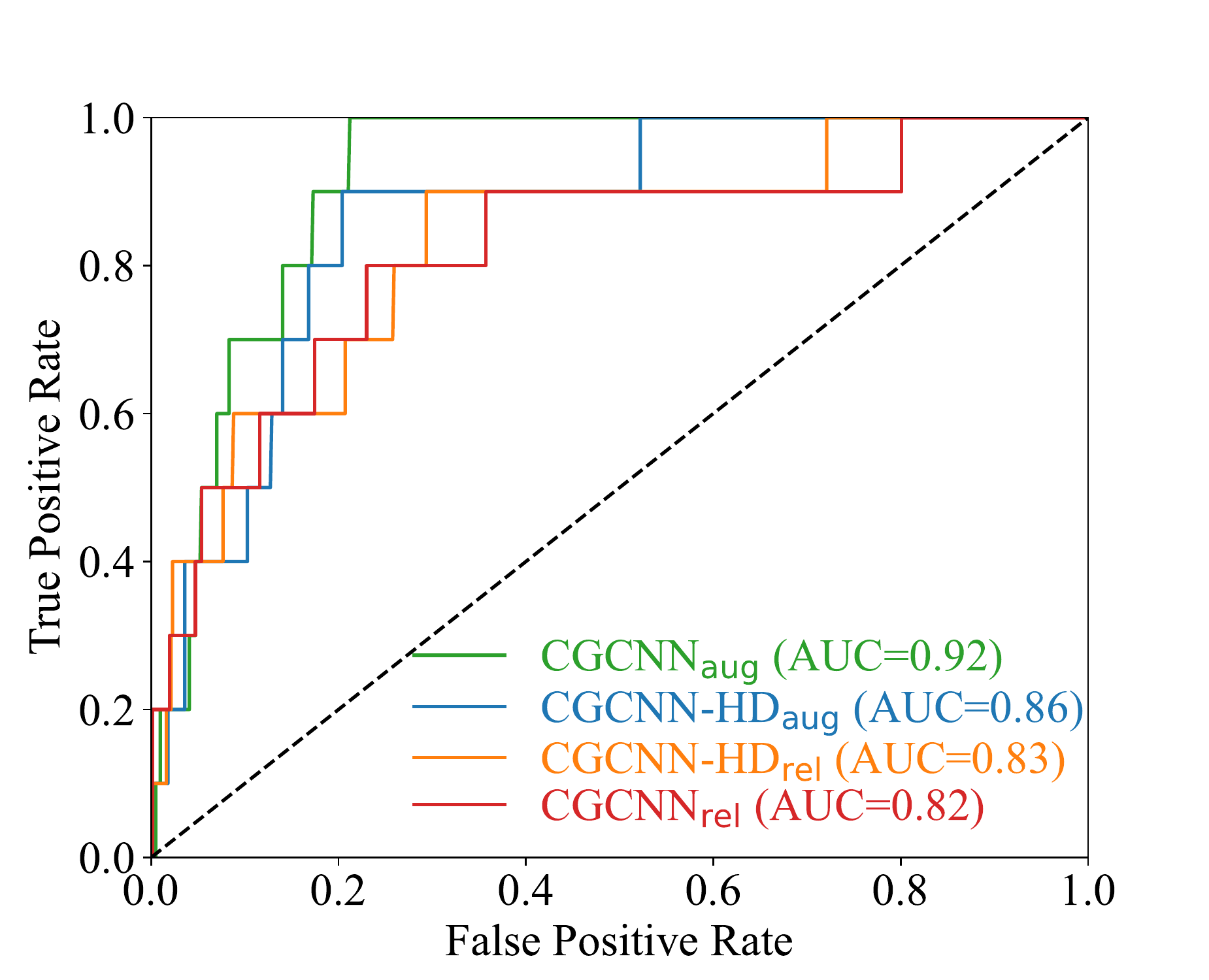}
    \caption{Receiver operating characteristic curve for the CGCNN and CGCNN-HD. the $aug$ subscript represents the model was trained on the augmented data. the $rel$ subscript represents the model was trained on the relaxed data. The dashed-black line represents a random classifier. The AUC is reported for all models.}
    \label{ROC}
\end{figure}

To construct the receiver operating characteristic (ROC) curve, shown in Figure~\ref{ROC} the predicted formation energies was used to compute hull distance ($\mathrm{E_{Hull}^{ML}}$). To compute a range of true positives, false positives, true negatives, and false negatives, we varied the stability criteria of $\mathrm{E_{Hull}^{ML}}$ over a range of hull distances that ensure a completed ROC curve for. We defined a true positive as a stable structure predicted as stable, a false positive as an unstable structure predicted as stable, a true negative as an unstable structure predicted as unstable, and a false negative as a  stable structure predicted as unstable. 

To further evaluate the models, we assume our test data is randomly generated and consider two hypothetical cases in which the models may be utilized. Case~1 emulates a study where identifying all stable structures is desirable (TPR = 1.0). Case~2  emulates a study where missing some stable structures is acceptable (TPR = 0.7). 

For case~1, the CGCNN trained on the augmented data performed best, successfully identifying all stable structures at a filtration criteria of $\mathrm{E_{Hull}^{ML} < 39~meV/atom}$ while misclassifying 130 structures,  yielding a 5-fold reduction in the number of energy calculations needed to identify all stable structures.  The CGCNN-HD trained on the augmented data misclassified 320 structures with a filtration criteria of $\mathrm{E_{Hull}^{ML} < 85~meV/atom}$. The CGCNN-HD trained on relaxed data misclassified 442 structures with optimal filtration criteria of $\mathrm{E_{Hull}^{ML} < 328~meV/atom}$. The CGCNN trained on relaxed data misclassified 491 structures with optimal filtration criteria of $\mathrm{E_{Hull}^{ML} < 576~meV/atom}$.

For case~2, again, the CGCNN trained on the augmented data performed best, obtaining a TPR of 0.7 and only misclassifying 54 structures with a filtration criteria of $\mathrm{E_{Hull}^{ML} < 6~meV/atom}$.  The CGCNN-HD trained on augmented data misclassified 89 structures with a filtration criteria of $\mathrm{E_{Hull}^{ML} < 21~meV/atom}$. The CGCNN trained on relaxed data misclassified 110 structures with a filtration criteria of $\mathrm{E_{Hull}^{ML} < 159~meV/atom}$.  The CGCNN-HD trained on relaxed data misclassified 130 structures a filtration criteria of $\mathrm{E_{Hull}^{ML} < 91~meV/atom}$.

The naive approach of setting the stability criteria to be the same for $\mathrm{E_{Hull}^{ML}}$ and $\mathrm{E_{Hull}^{DFT}}$, restricts the ability to select a balance of accuracy and computational cost. for example at stability criteria of $\mathrm{E_{Hull}^{ML} < 0}$, the CGCNN trained with the augmented dataset obtained a TPR of 0.6, misclassifying  47 structures. While this performance is acceptable, as shown previously, at stability criteria of $\mathrm{E_{Hull}^{ML} < 39~meV/atom}$ the model can correctly identify all stable structures which may be more desirable for a given application.

\section{Conclusion}
We proposed a simple, physically motivated, computationally cheap perturbation technique that augmented our data to better represent the PES, dramatically improving unrelaxed structure formation energy predictions.  When compared to training on only relaxed structures, training with an augmented data set consisting of one relaxed and one perturbed structure for every relaxed structure, prediction MAEs of the CGCNN and CGCNN-HD was reduced from 251~meV/atom and 172~meV/atom to 86~meV/atom and 82~meV/atom, respectively. Further, we showed that predictions on relaxed structures showed an anti-correlation to predictions of unrelaxed structures. We then showed that augmenting the data enabled the predictions of relaxed structures to follow a similar trend to that of unrelaxed structures. Finally, we utilize a ROC curve to show two cases where our method may be useful in accelerating CSPA.  While there likely exist more advanced augmentation techniques, this work showed the surprising effectiveness of a relatively simple method of augmentations, that outperformed the current state of the art in formation energy prediction of unrelaxed structures. 

\section{Code Availability}
Code for implementing the model on both cpus and gpus, training the models, augmenting training data are available at \url{https://github.com/JasonGibsonUfl/Augmented_CGCNN}. Data will be made available upon reasonable requests.
\section{Acknowledgments}
This work was supported by the National Science Foundation under grants Nos. PHY-1549132, the Center for Bright Beams,  and  the software fellowship awarded to J.B.G. by the Molecular Sciences Software Institute funded by the National Science Foundation (Grant No. ACI-1547580). Computational resources were provided by the University of Florida Research Computing Center. 
\section{Author Contributions}
JBG and RGH conceived the Augmentation technique and training strategy. JBG implemented the Augmentation technique and performed the model training and analysis. ACH performed the genetic algorithm structure search to provide testing data. JBG, ACH, and RGH contributed to the writing of the manuscript.

\bibliography{ref}

\end{document}